\documentclass[final,5p,times,twocolumn]{elsarticle}
\usepackage{amssymb}
\usepackage{amsmath}
\usepackage{color}
\usepackage{amsthm}
\usepackage{dcolumn}
\usepackage{lineno}
\usepackage{cuted}

\newenvironment{widetext}
{\begin{strip}}
{\end{strip}}

\journal{Chaos, Solitons and Fractals}

\begin{document}

\begin{frontmatter}

\title{Vibrational resonance in a frequency-adaptive learning Duffing system}

\author[label1]{Zhongqiu Wang}
\address[label1]{School of Computer Science and Technology, China University of Mining and Technology, Xuzhou 221116, Jiangsu, People's Republic of China}

\author[label2]{Jianhua Yang\corref{cor1}}
\ead{jianhuayang@cumt.edu.cn}
\cortext[cor1]{Corresponding author}

\author[label2]{Feng Tian}

\address[label2]{School of Mechatronic Engineering, China University of Mining and Technology, Xuzhou, 221116, Jiangsu, People's Republic of China}

\author[label3]{Huatao Chen}
\address[label3]{Division of Dynamics and Control, School of Mathematics and Statistics, Shandong University of Technology, Zibo 255000, People's Republic of China}

\author[label4,label5]{Miguel A. F. Sanju\'an}
\address[label4]{Nonlinear Dynamics, Chaos and Complex Systems Group, Departamento de Fisica, Universidad Rey Juan Carlos, Tulipan s/n, Mostoles, 28933, Madrid, Spain}
\address[label5]{Royal Academy of Sciences of Spain, Valverde 22, 28004 Madrid, Spain}
\begin{abstract}
Vibrational resonance focuses on the resonance behavior of a nonlinear system when it is subjected to both a weak low-frequency characteristic signal and a high-frequency auxiliary signal. A traditional Duffing system has a fixed natural frequency and lacks adaptability to the excitation frequency, resulting in vibrational resonance occurring only in a lower frequency range, which affects the application of vibrational resonance. We propose a frequency-adaptive learning Duffing system to overcome the above problem through a learning rule of the natural frequency. The optimal vibrational resonance performance is demonstrated by examining the influence of auxiliary signal parameters, nonlinear stiffness coefficient and the learning rule on the response. The appearance of vibrational resonance is verified by numerical simulation, approximated theoretical predication and circuit simulation. In addition, the advantages of the proposed frequency-adaptive learning rule are highlighted in vibrational resonance performance by comparing with that of two other commonly used alternatives called Hebbian learning rules. The proposed learning rule makes the system more stable and have a stronger resonance degree. The results provide a useful reference for optimizing nonlinear system response and also for processing a weak characteristic signal through nonlinear resonance methods. These achievements provide a groundbreaking foundation for future applied studies especially in the field of weak and complex signal processing.
\end{abstract}
\end{frontmatter}



\section{Introduction}
The Duffing oscillator is a typical type of nonlinear oscillator and one of the most widely studied nonlinear systems \cite{Ref1}. It has broad applications in modeling nonlinear vibration and structural dynamics across multiple disciplines. In mechanical engineering, the Duffing oscillator is used to describe phenomena such as vibrations during machine tool processing \cite{Ref2}, the dynamic response of automotive suspension systems \cite{Ref3}, vibration control of robotic arms \cite{Ref4}, large-deflection vibrations of beams or plates \cite{Ref5}, and vibrations in gear and bearing systems \cite{Ref6}. In acoustics, it is employed to model the dynamics of nonlinear acoustic components, including nonlinear acoustic filters \cite{Ref7} and ultrasonic vibration systems \cite{Ref8}. In electrical engineering, Duffing-type dynamics arise in nonlinear circuits and electromagnetic vibration systems, where nonlinear components (e.g., inductors, capacitors, semiconductor devices) or electromagnetic coupling effects lead to such behavior. Typical examples include nonlinear RLC circuits \cite{Ref9}, vibrations in motors and transformers \cite{Ref10}, and piezoelectric vibration energy harvesters \cite{Ref10}. The Duffing oscillator also plays a role in civil engineering, capturing the nonlinear response of structures such as the seismic response of buildings \cite{Ref11}, bridge vibrations induced by wind or vehicles \cite{Ref12}, and wave responses of offshore platforms \cite{Ref13}. Beyond engineering, it has applications in biology and medicine, where it is used to describe nonlinear physiological vibrations, including those in the cardiovascular system \cite{Ref14}, nonlinear auditory responses \cite{Ref15}, and cellular or molecular vibrations \cite{Ref16}. Moreover, Duffing systems have found significant use in quantum physics \cite{Ref17} and astrophysics \cite{Ref18}. In addition to the examples above, numerous nonlinear system models across diverse fields can be effectively described by Duffing equations. Accordingly, the Duffing system has important theoretical significance and engineering application values.\\
\indent The Duffing system exhibits a wide range of rich nonlinear dynamical behaviors, including chaos \cite{Ref19}, bifurcation \cite{Ref20}, fractals \cite{Ref21}, and various complex resonances \cite{Ref22, Ref23}. This work focuses on vibrational resonance, especially on enhancing a weak characteristic component in the output by adjusting an auxiliary signal with much higher frequency. In this way, the cooperation between the nonlinear system and the auxiliary signal can amplify the weak characteristic signal greatly \cite{Ref24, Ref25}. Early research on vibrational resonance was extensively carried out using the Duffing system \cite{Ref26, Ref27, Ref28}. The present study differs in that it addresses how to induce vibrational resonance when the characteristic frequency is in a much higher scope. Existing approaches, such as the re-scaling method \cite{Ref29, Ref30} and the twice-sampling method \cite{Ref31}, can achieve high-frequency signal processing effect but require demanding parameter tuning to obtain satisfactory resonance performance \cite{Ref32, Ref33}. To overcome this limitation, this work draws on the concept of frequency-adaptive oscillator \cite{Ref34, Ref35, Ref36, Ref37, Ref38}. An effective frequency-adaptive rule can enable the system to resonate across a broader frequency range \cite{Ref39, Ref40, Ref41, Ref42, Ref43}. The frequency-adaptive learning systems are currently at the forefront of research and have been studied and applied in various disciplines such as reservoir computer \cite{Ref44}, microgrid system \cite{Ref45}, assistive wearable device \cite{Ref46}, and so on. The key to design a frequency-adaptive learning system is to have an appropriate learning rule that enables the natural frequency of the system to better track the change in the external excitation. Based on this idea, a new frequency-adaptive learning Duffing system is proposed here, incorporating a novel adaptive-frequency learning rule. With this rule, the system demonstrates significantly stronger vibrational resonance in a wide frequency range. The characteristic of the proposed adaptive-frequency learning Duffing system on vibrational resonance is revealed not only by numerical study but also by an approximated theoretical predication, and the output of the simulation circuit which is also widely used in the system performance verification \cite{Ref47, Ref48, Ref49}.\\
\indent The structure of this work is as follows. Section~2 examines the vibrational resonance phenomenon in the conventional monostable Duffing system, serving as a baseline for comparison with the proposed approach and highlighting the advantages of vibrational resonance in the new system. Section~3 introduces the frequency-adaptive learning Duffing system and demonstrates how the proposed learning rule facilitates vibrational resonance. Section~4 builds a circuit corresponding to the frequency-adaptive learning Duffing system to display the existence of vibrational resonance in another way. Section~5 gives a comparative analysis with two other commonly used Hebbian learning rules further illustrates the superior performance of the proposed learning rule. It also clarifies the limitation of the proposed system and the prospect of the work. Finally, Section~6 summarizes the main conclusions of this study.
\section{Vibrational resonance in the conventional Duffing system}
Considering a conventional dimensionless monostable Duffing system that excited by two harmonic signals simultaneously, the governed equation is
\begin{equation}
\ddot x + 2\zeta {\omega _0}\dot x + {\omega _0}^2x + b{x^3} = A\cos ({\Omega _1}t) + B\cos ({\Omega _2}t).
\end{equation}
In Eq.~(1), $\zeta$ is the damping ratio and $\omega _0$ is the natural frequency of the corresponding linear system. The parameter $b$ is the nonlinear stiffness coefficient which can be a positive value or a negative value. The stiffness term represents as a harden spring for the case $b>0$ but as a soften spring for the case $b<0$. The frequencies of the two harmonic signals satisfy $\Omega_1 \ll \Omega_2$. We refer to $A\cos ({\Omega _1}t)$ as the characteristic signal, and $B\cos ({\Omega _2}t)$ as the auxiliary signal which is used to induce vibrational resonance.\\
\indent This work mainly studies the phenomenon of vibrational resonance, which is measured by the response amplitude $Q$, and the numerical calculation formula for $Q$ is
\begin{equation}
Q = \frac{{\sqrt {{Q_s}^2 + {Q_c}^2} }}{A}.
\end{equation}
In Eq.~(2), $Q_s$ and $Q_c$ are the sine and cosine Fourier coefficients computed through the time series, respectively. The response amplitude $Q$ indicates the amplification of the weak low-frequency signal after it passes through the nonlinear system. As a result, the sine and cosine Fourier coefficients should also be extracted at the characteristic frequency $\Omega_1$, i.e.,
\begin{equation}
\left\{ \begin{array}{l}
 {Q_{\rm{s}}} = \frac{2}{{mT}}\int\limits_0^{mT} x (t)\sin ({\Omega _1}t){\rm{d}}t \\
 {Q_{\rm{c}}} = \frac{2}{{mT}}\int\limits_0^{mT} x (t)\cos ({\Omega _1}t){\rm{d}} \\
 \end{array} \right.
\end{equation}
They can be computed from discrete time series of $x(t)$ by the following expressions
\begin{equation}
\left\{ \begin{array}{l}
 {Q_{\rm{s}}} = \frac{{2\Delta t}}{{mT}}\sum\limits_{i = 1}^{mT/\Delta t} x ({t_i})\sin ({\Omega _1}{t_i}) \\
 {Q_{\rm{c}}} = \frac{{2\Delta t}}{{mT}}\sum\limits_{i = 1}^{mT/\Delta t} x ({t_i})\cos ({\Omega _1}{t_i}) \\
 \end{array} \right.
\end{equation}
In Eq.(3) and Eq.(4), $mT$ is the length of time for calculating the Fourier coefficients $Q_{\rm{s}}$ and $Q_{\rm{s}}$, where $m$ is usually a positive integer and $T=2\pi/\Omega_1$ is the period of the characteristic signal. In addition, $m$ should be large enough and obtained from the steady response. In the numerical simulation of this work, we let $m=200$. Specifically, we get the time series of $x(t)$ for the time length of $250T$ at first, and then we remove the time series of the first $50T$ to obtain the so-called steady response. In other words, the time series of the last $200T$ time length is used to calculate the response amplitude $Q$. The time step $\Delta t=0.001$ and the initial conditions $x(0) = 0$ and $\dot x(0) = 0$ are selected, and an Euler algorithm is applied in the following numerical calculations.\\
\indent Figure 1 shows the functional relationship between the response amplitude $Q$ calculated according to Eq.~(2) and the auxiliary signal amplitude $B$. From the curves in Fig.~1(a), although a resonance-like phenomenon occurs for the case $b>0$, it is not a bona fide resonance because the value of $Q$ is relatively small, slightly greater than 1, which indicates that the amplitude of the characteristic signal is almost not amplified through the action of the nonlinear system and the auxiliary signal. In Fig.~1(b), the value of $Q$ becomes very small for the case $b<0$. In this case, the characteristic signal is not only amplified but also the response of the system tends to diverge as $B$ increasing. This behavior is related to the potential function of the system, which exhibits a double-hump single-well structure, making the response easily prone to divergence.
\begin{figure}
\includegraphics[width=0.4 \textwidth]{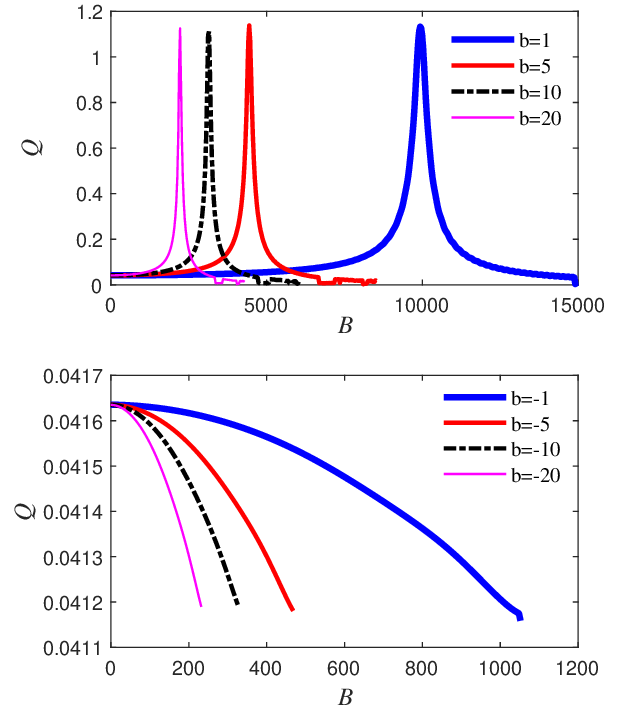}
\caption{The response amplitude $Q$ versus the high-frequency signal amplitude $B$ and the nonlinear stiffness coefficient $b$ calculated from Eq.~(1). (a) There is not a bona fide vibrational resonance due to the small value of $Q$ when $b>0$. (b) There is not vibrational resonance appearing and the response diverges as $B$ increasing when $b<0$. The simulation parameters are $\zeta=0.1$, $\omega_0=1$, $A=0.1$, $\Omega_1=5$ and $\Omega_2=50$.}
\end{figure}
\\
\indent From a signal-processing perspective, the excitation in Eq.~(1) is amplified by a factor of $\beta$ before being applied to the nonlinear system. At this point, Eq.~(1) changes to
\begin{equation}
\ddot x + 2\zeta {\omega _0}\dot x + {\omega _0}^2x + b{x^3} = \beta [ A\cos ({\Omega _1}t) + B\cos ({\Omega _2}t)].
\end{equation}
\indent We simulate the response of Eq.~(5) and obtain the relationship between the response amplitude $Q$ and the auxiliary signal amplitude $B$, as shown in Fig.~2. When $b$ takes a positive value, the system response exhibits vibrational resonance. The critical value $B_c$ and the maximal value $Q_{max}$ corresponds to the resonance peak depend on the value of $b$. However, when $b$ takes a negative value, the response does not undergo vibrational resonance and is easily prone to divergence.
\begin{figure}
\includegraphics[width=0.4 \textwidth]{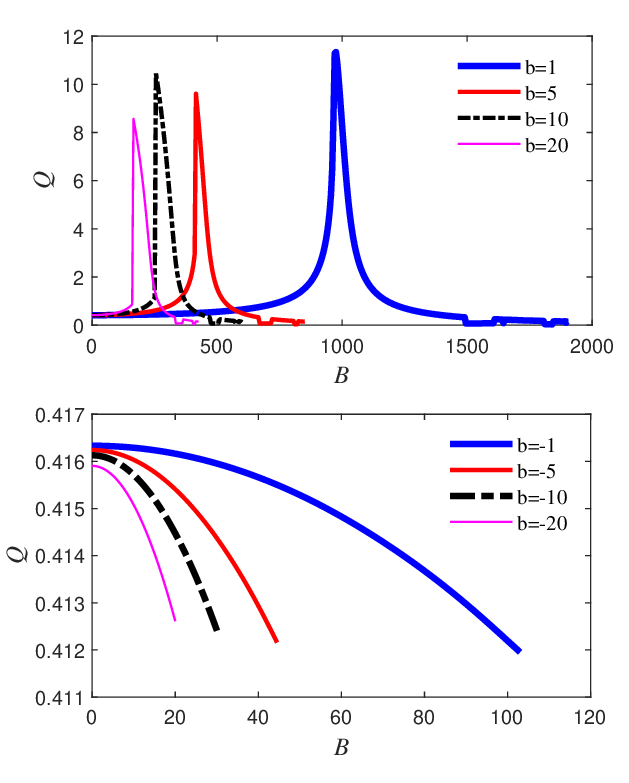}
\caption{The response amplitude $Q$ versus the high-frequency signal amplitude $B$ and the nonlinear stiffness coefficient $b$ calculated from Eq.~(5). (a) Vibrational resonance appears when $b>0$ indicating the necessity of the amplification $\beta$. (b) There is not vibrational resonance appearing and the response diverges as $B$ increasing when $b<0$. The simulation parameters are $\zeta=0.1$, $\omega_0=1$, $\beta=10$, $A=0.1$, $\Omega_1=5$ and $\Omega_2=50$.}
\end{figure}
\\
\indent Therefore, the results in Figs.~1 and 2 indicate that when $\Omega_1$ is large, inducing vibrational resonance in the output requires choosing a positive value for $b$ and amplifying both the characteristic and auxiliary signals by a factor $\beta$ before applying them to the nonlinear system.\\
\indent In fact, the results in Figs.~1 and 2 can also be validated analytically. The method of direct separation of fast and slow motions provides a simple yet effective approach for obtaining the analytical response amplitude \cite{Ref24, Ref28, Ref50, Ref51, Ref52}. However, since extensive studies on this method already exist in the cited works and analytical study is not the primary focus of this paper, hence it will not be detailed here.
\section{Vibrational resonance in the frequency-adaptive learning Duffing system}
\subsection{A modified Duffing system with a frequency-adaptive learning rule}
We further modify Eq.~(5) so that the natural frequency of the Duffing system adapts the external excitations under the role of a learning function. The modified Duffing system is given by
\begin{equation}
\left\{ \begin{array}{l}
 \ddot x + 2\zeta \dot x + {\omega ^2}x + b{x^3} = \beta [ A\cos ({\Omega _1}t) + B\cos ({\Omega _2}t)] \\
 \dot \omega  = {k_\omega }[A\cos ({\Omega _1}t) + B\cos ({\Omega _2}t)] \\
 \end{array} \right.
\end{equation}
It is the so-called frequency-adaptive learning Duffing system. Apparently, the fixed natural frequency $\omega_0$ in Eq.~(5) is replaced by a simple nonlinear function $\omega$. Specifically, the nonlinear function $\omega$ can be solved by $\dot \omega$ which is called the learning rule, and the symbol $k_\omega$ is the learning rate. Edmon Perkins \cite{Ref41, Ref42} proposed this kind of frequency-adaptive system with a learning rule and conducted research on it. We are mainly inspired by his two works to study the vibrational resonance phenomenon in the adaptive learning system. The learning rule plays a dominant role in the response of a nonlinear system. Herein, we give a different learning rule and will compare it with the learning rule that called Hebbian learning rule used by Edmon Perkins later. Moreover, in Eq.~(6), the natural frequency is removed from the damping term, it avoids further complicating the research theme. This handling way is also adopted by the former works by Edmon Perkins pointed out aforementioned. Another important reason for removing the natural frequency from the damping term is that the nonlinear system studied in this work is not specific to a particular physical model, but rather propose a class of nonlinear systems whose main function is for weak signal processing. In this regard, the system in Eq.~(6) meets our research theme, so this work removes the natural frequency from the damping term when studying the vibrational resonance under the role of learning rule.\\
\indent In Eq.~(6), integrating $\omega$ over the time $t$ yields
\begin{equation}
\omega  = \frac{{A{k_\omega }}}{{{\Omega _1}}}\sin ({\Omega _1}t) + \frac{{B{k_\omega }}}{{{\Omega _2}}}\sin ({\Omega _2}t)
\end{equation}
From the expression of $\omega$ in Eq.~(7), the natural frequency of the modified Duffing system is directly related to the external excitation. Then, it is easy to obtain
\begin{equation}
{\omega ^2} = \frac{{{k_\omega }^2}}{2}\left( {\frac{{{A^2}}}{{{\Omega _1}^2}} + \frac{{{B^2}}}{{{\Omega _2}^2}}} \right) - \frac{{{k_\omega }^2{A^2}}}{{2{\Omega _1}^2}}\cos (2{\Omega _1}t) - \frac{{{k_\omega }^2{B^2}}}{{2{\Omega _2}^2}}\cos (2{\Omega _2}t)
\end{equation}
Obviously, the positive and negative values of $k_\omega$ will not affect the response, so we let $k_\omega > 0$ in the following analysis. Substituting the expression of $\omega^2$ into Eq.~(6), we get the concrete expression for the modified Duffing system is

\begin{widetext}
\begin{equation}
\ddot x + 2\zeta \dot x + \frac{{{k_\omega }^2}}{2}\left( {\frac{{{A^2}}}{{{\Omega _1}^2}} + \frac{{{B^2}}}{{{\Omega _2}^2}}} \right)x - \frac{{{k_\omega }^2{A^2}}}{{2{\Omega _1}^2}}\cos (2{\Omega _1}t)x - \frac{{{k_\omega }^2{B^2}}}{{2{\Omega _2}^2}}\cos (2{\Omega _2}t)x + b{x^3} = \beta [A\cos ({\Omega _1}t) + B\cos ({\Omega _2}t)]
\end{equation}
\end{widetext}
\indent On the one hand, for Eq.~(9), because the external excitations and the parametric excitations exist at the same time, and there are four explicit frequencies in the excitation, the general method of direct separation of fast and slow motions is no longer applicable here. The aim of this paper is to study the vibrational resonance and reveal the new impact of frequency adaptation on the phenomenon. Hence, the following analysis of this section will mainly focus on numerical calculations.\\
\indent On the other hand, we can perform some approximated analysis on the resonant frequency through Eq.~(9). Apparently, the natural frequency of the above system is determined by the expression of $\omega^2$. Assuming the constant component of $\omega^2$ dominates the natural frequency of the system, the vibrational resonance condition at the excitation frequency $\Omega_1$ is $\Omega_1=\omega$. Hence, we have
\begin{equation}
{\Omega _1}^2 = \frac{{{k_\omega }^2}}{2}\left( {\frac{{{A^2}}}{{{\Omega _1}^2}} + \frac{{{B^2}}}{{{\Omega _2}^2}}} \right)
\end{equation}
Under this assumption, we calculate the critical value $B_c$ which corresponds to the resonance peak and get
\begin{equation}
{B_c} = \sqrt {\frac{{2{\Omega _2}^2}}{{{k_\omega }^2}}\left( {{\Omega _1}^2 - \frac{{{k_\omega }^2{A^2}}}{{2{\Omega _1}^2}}} \right)}.
\end{equation}
If we make $\Omega_2$ as the controllable parameter, the critical value $\Omega_{2c}$ solving by the vibrational resonance condition of Eq.~(10) is
\begin{equation}
{\Omega _{2c}} = \frac{{{k_\omega }B{\Omega _1}}}{{\sqrt {2{\Omega _1}^4 - {k_\omega }^2{A^2}} }}
\end{equation}
Similarly, we can also get the critical excitation frequency $\Omega _{1c}$ corresponding to the resonance peak is
\begin{equation}
{\Omega _{1c}} = \sqrt {{{\left( {\frac{{{k_\omega }B}}{{2{\Omega _2}}}} \right)}^2} + \sqrt {{{\left( {\frac{{{k_\omega }B}}{{2{\Omega _2}}}} \right)}^4} + \frac{1}{2}{{({k_\omega }A)}^2}} }
\end{equation}
Further, if we know $\Omega _2/\Omega _1 = \gamma$ , then from Eq.~(10) we obtain the other expression for $\Omega _{1c}$ is
\begin{equation}
{\Omega _{1c}} = \sqrt[4]{{\frac{{{k_\omega }^2}}{4}\left( {{A^2} + \frac{{{B^2}}}{{{\gamma ^2}}}} \right)}}
\end{equation}
\indent Herein, we must emphasize that this assumption of resonance analysis is only an approximation. The resonance of a nonlinear system is a very complicated problem, and the nonlinear terms such as parametric excitations and the nonlinear stiffness term still affect the vibrational resonance of the system in some cases. Moreover, compared to the traditional Duffing system, the modified Duffing system under the influence of the frequency-adaptive learning rule has become more complex, and it may lead to richer dynamic phenomena in the response. Considering the structure and theme of the work, this work focuses on the numerical results of vibrational resonance and reveal the new impact of frequency adaptation on vibrational resonance.
\subsection{Vibrational resonance induced by the auxiliary signal}
\indent Now, we study the occurrence of vibrational resonance of Eq.~(6), considering the vibrational resonance caused by the auxiliary signal. From Fig.~3(a), it can be seen that there is obvious vibrational resonance appearing, and the critical value $B_c$ at which the resonance occurs, is relatively small compared to that of Fig.~2(a). Consequently, the vibrational resonance behavior in Fig.~3 differs significantly from that observed in Figs.~1 and 2. It is completely feasible to achieve a bona fide vibrational resonance using the proposed frequency-adaptive Duffing system in Eq.~(6). The relationship between the maximum response amplitude $Q_{max}$ and the corresponding critical value $B_c$ is shown in Fig.~3(b), and there is a certain fluctuation between them. In Fig.~3(c), the numerical results of $B_c$ versus the frequency $\Omega_2$ are basically consistent with the approximated theoretical results that given in Eq.~(11). It indicates that the constant term of $\omega^2$ in Eq.(8) indeed plays a decisive role in the natural frequency of the system of Eq.(9). In addition, the constant term of $\omega^2$ is generated due to the presence of excitations in the learning rule. Consequently, the learning rule results in the vibrational resonance is related to the auxiliary signal directly. In Fig.~3(d), the relationship between $\Omega_2$ and $Q_{max}$ presents fluctuating in similar to that in Fig.~3(b). It should be noted that the white area in Fig.~3 is the area of the response divergence. The white areas in the following figures also have the same meaning.
\begin{figure}
\center
\includegraphics[width=0.5 \textwidth]{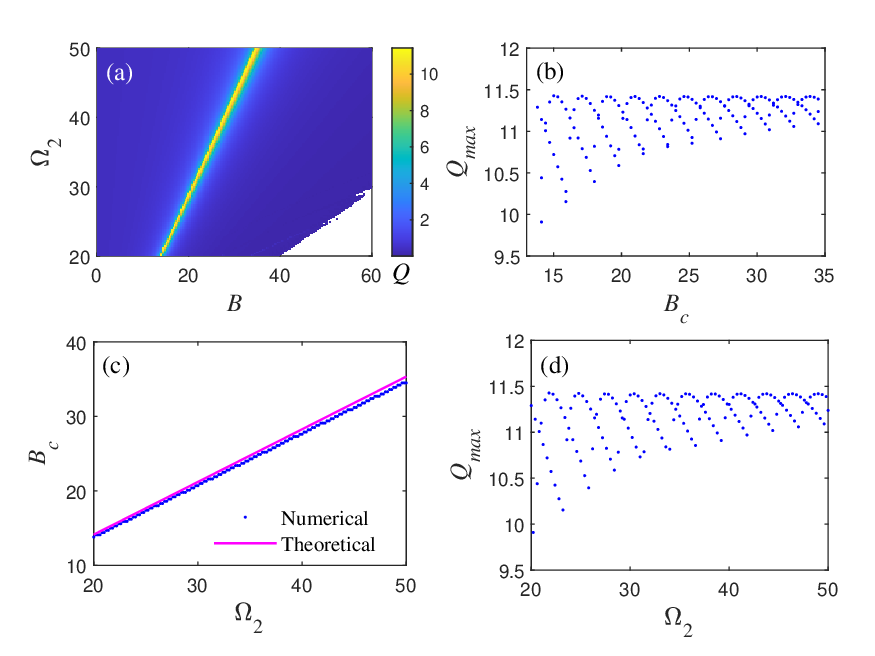}
\caption{Vibrational resonance induced by the auxiliary signal. (a) The response amplitude $Q$ presents obvious vibrational resonance band in the $B-\Omega_2$ plane. (b) The maximal value of the response amplitude $Q_{max}$ versus the corresponding critical value $B_c$. (c) The critical values of $B_c$ versus the high-frequency $\Omega_2$ are obtained by numerical simulation and approximated theoretical predication in Eq.~(11) respectively. (d) The maximal value of the response amplitude $Q_{max}$ versus the high-frequency $\Omega_2$. The simulation parameters are $\zeta=0.1$, $b=1$, $k_\omega=10$, $\beta=10$, $A=0.1$ and $\Omega_1=5$.}
\end{figure}
\\
\indent In Fig.~4, we investigate the effect of the excitation amplification factor $\beta$ on vibrational resonance that caused by the signal amplitude $B$. As shown in Figs.~4(a) and 4(d), increasing $\beta$ leads to a larger value of $Q_{max}$. Within a certain range, the parameter $\beta$ has a relatively small impact on the critical value of $B_c$, as shown in Figs.~4(b) and 4(c). A reasonable value of $\beta$ can achieve a satisfactory vibrational resonance performance.
\begin{figure}
\center
\includegraphics[width=0.5 \textwidth]{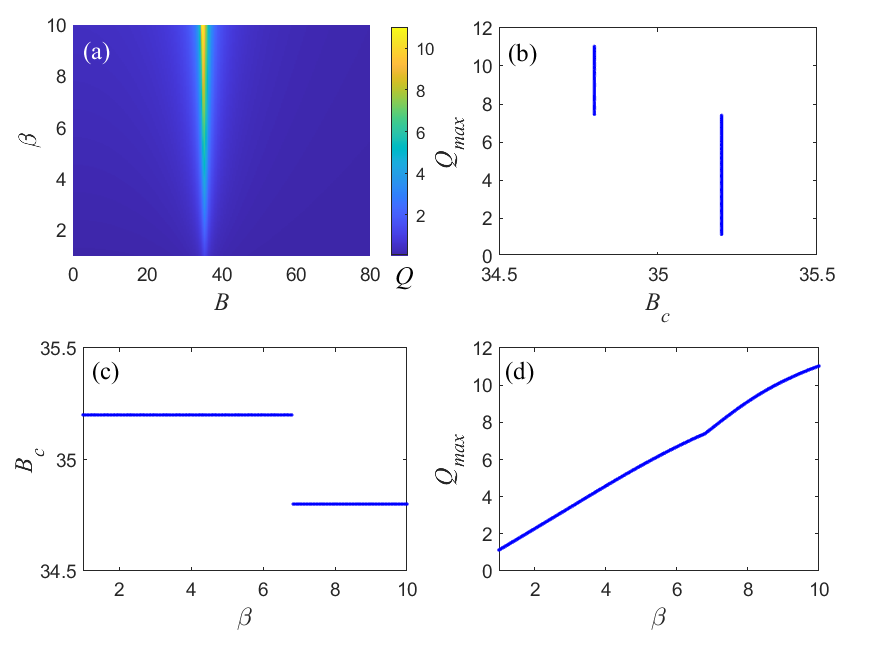}
\caption{Effect of the amplification factor $\beta$ on vibrational resonance that induced by the auxiliary signal. (a) The response amplitude $Q$ presents obvious vibrational resonance band in the $B-\beta$ plane. (b) The maximal value of the response amplitude $Q_{max}$ versus the corresponding critical value $B_c$. (c) Critical value $B_c$ versus the excitation amplification factor $\beta$. (d) The maximal value of the response amplitude $Q_{max}$ versus the amplification factor $\beta$. The simulation parameters are $\zeta=0.1$, $b=1$, $k_\omega=10$, $A=0.1$, $\Omega_1=5$ and $\Omega_2=50$.}
\end{figure}
\\
\indent In Fig.~5, we investigate the occurrence of vibrational resonance in the system as the characteristic frequency $\Omega_1$ increasing. Figure 5(a) shows that a higher $\Omega_1$ requires a larger value of $B$ for the response to exhibit vibrational resonance. The smaller the value of $\Omega_1$, the larger the maximal value of the response amplitude $Q_{max}$ in the curve. As seen in Figs.~5(a) and 5(b), when $B = 0$ the system undergoes strong resonance driven solely by the characteristic signal, but this occurs only when $\Omega_1$ is small. For larger $\Omega_1$, the auxiliary signal becomes necessary to induce strong resonance. The critical value $B_c$ that causes resonance is shown in Fig.~5(b), which also shows the fluctuation between the critical value $B_c$ and the maximum value of the response amplitude $Q_{max}$ at the resonance peak. Figure 5(c) illustrates a nonlinear growth relationship between the critical value $B_c$ and the characteristic frequency $\Omega_1$ by both the numerical simulation results and the approximated theoretical predication in Eq.~(11). Figure 5(d) reflects a certain fluctuation relationship between the characteristic frequency $\Omega_1$ and the maximum response amplitude $Q_{max}$, which is consistent with the facts reflected in Figs.~5(b) and 5(c). In summary, Fig.~5 demonstrates that as $\Omega_1$ increases, adjusting the amplitude of the auxiliary signal can still induce vibrational resonance in the system.
\begin{figure}
\center
\includegraphics[width=0.5 \textwidth]{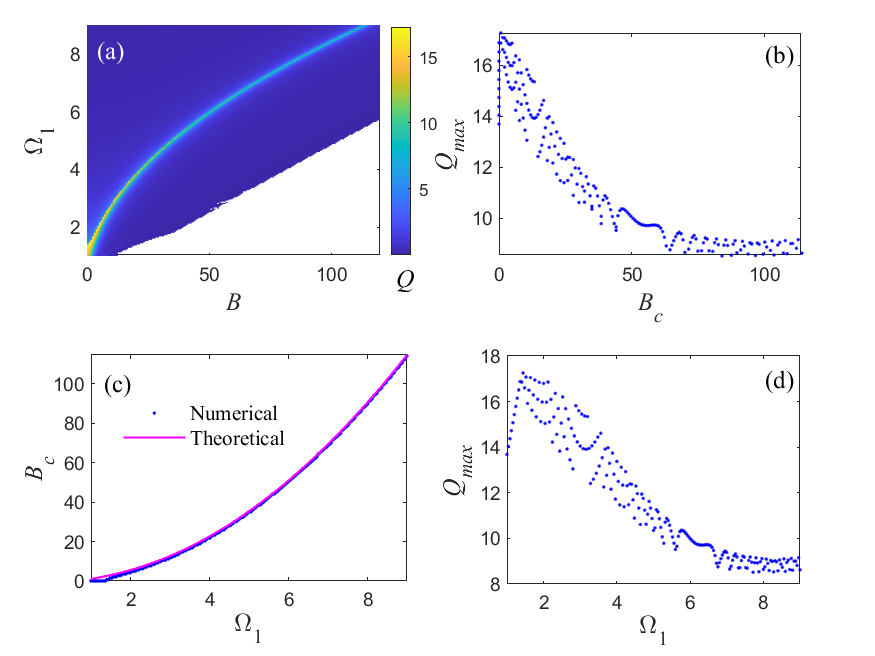}
\caption{Effect of the characteristic frequency $\Omega_1$ on vibrational resonance that induced by the auxiliary signal. (a) The response amplitude $Q$ presents obvious vibrational resonance band in the $B-\Omega_1$ plane. (b) The maximal value of the response amplitude $Q_{max}$ versus the corresponding critical value $B_c$. (c) The critical values of $B_c$ versus the characteristic frequency $\Omega_1$ are obtained by numerical simulation and approximated theoretical predication in Eq.~(11) respectively. (d) The maximal value of the response amplitude $Q_{max}$ versus the signal frequency $\Omega_1$. The simulation parameters are $\zeta=0.1$, $b=1$, $k_\omega=10$, $\beta=10$, $A=0.1$ and $\Omega_2 = 10\Omega_1$.}
\end{figure}
\subsection{Effect of the nonlinear stiffness coefficient on vibrational resonance}
\indent In Fig.~6, we consider the effect of the nonlinear stiffness coeffcient $b$ on vibrational resonance. As shown in this figure, regardless of $b$ taking a positive or negative value, the system will experience vibrational resonance. In contrast, when $b$ is negative in Figs.~1 and 2, the system will not experience vibrational resonance, and its response is also prone to divergence. It also demonstrates that the system proposed in Eq.~(6) has significant advantages over the conventional Duffing system in terms of promoting vibrational resonance and improving the stability of the system. Specifically, the role of the learning rule not only makes vibrational resonance be easily induced, but the stability of the system response is also improved over a large range. The relationship between the resonance peak $Q_{max}$ and the corresponding critical value $B_c$ is shown in Fig.~6(b). This graph illustrates the phenomenon of vibrational resonance, where a single $B_c$ value corresponds to multiple $Q_{max}$ values, which is also significantly different from vibrational resonance in the conventional monostable Duffing system. This phenomenon occurs because different values of $b$ can yield the same $B_c$, as shown in Fig.~6(c), while producing different $Q_{\max}$ values, as shown in Fig.~6(d).
\begin{figure}
\center
\includegraphics[width=0.5 \textwidth]{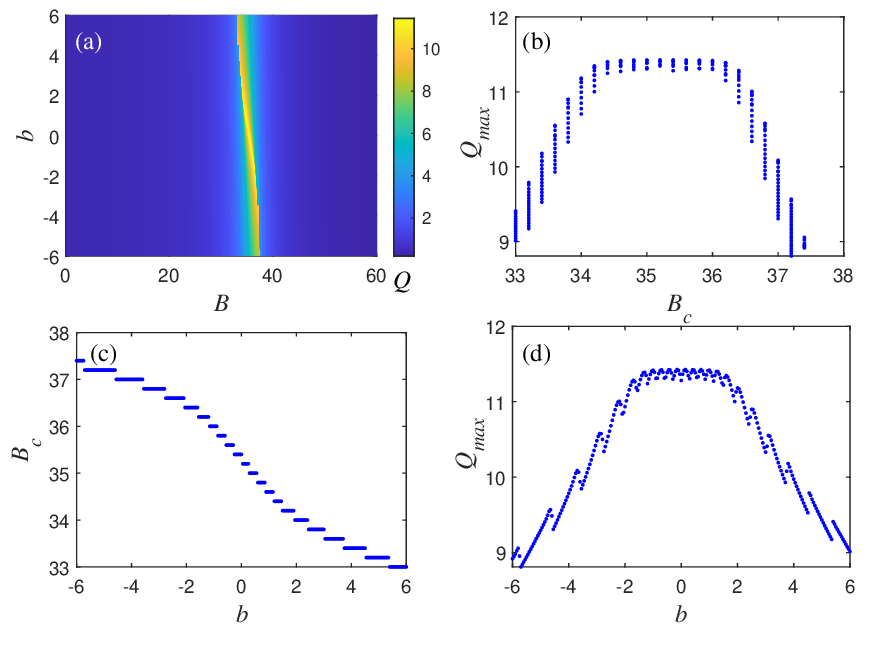}
\caption{Effect of the nonlinear stiffness coefficient $b$ on vibrational resonance that induced by the auxiliary signal. (a) The response amplitude $Q$ presents obvious vibrational resonance band in the $B-b$ plane. (b) The maximal value of the response amplitude $Q_{max}$ versus the corresponding critical value $B_c$. (c) Critical value $B_c$ versus the nonlinear stiffness coefficient $b$. (d) The maximal value of the response amplitude $Q_{max}$ versus the nonlinear stiffness coefficient $b$. The simulation parameters are $\zeta=0.1$, $k_\omega=10$, $\beta=10$, $A=0.1$, $\Omega_1 = 5$ and $\Omega_2 = 50$.}
\end{figure}
\\
\indent In Fig.~7, with the frequency $\Omega_2$ as the control parameter, vibrational resonance appears when $\Omega_2$ varies from negative to positive. The relationship between the maximal response amplitude $Q_{\max}$ and the corresponding critical value $B_c$ exhibits a pattern similar to that observed in Fig.~6.
\begin{figure}
\center
\includegraphics[width=0.5 \textwidth]{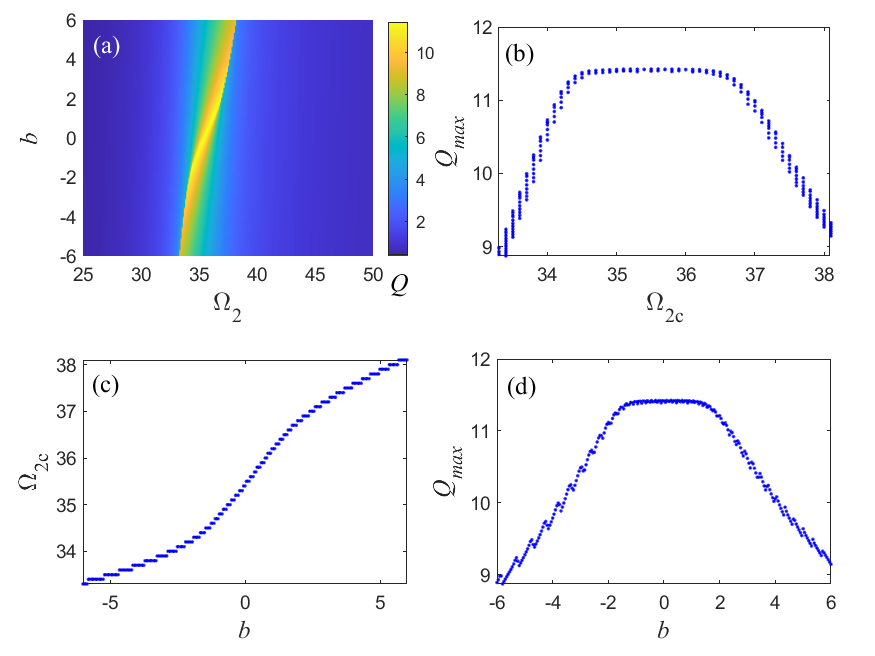}
\caption{Effect of the nonlinear stiffness coefficient $b$ on vibrational resonance that induced by the auxiliary signal. (a) The response amplitude $Q$ presents obvious vibrational resonance band in the $\Omega_2-b$ plane. (b) The maximal value of the response amplitude $Q_{max}$ versus the corresponding critical value $\Omega_{2c}$. (c) Critical value $\Omega_{2c}$ versus the the nonlinear stiffness coefficient $b$. (d) The maximal value of the response amplitude $Q_{max}$ versus the nonlinear stiffness coefficient $b$. The simulation parameters are $\zeta=0.1$, $k_\omega=10$, $\beta=10$, $A=0.1$, $\Omega_1=5$ and $B=25$.}
\end{figure}
\subsection{Effect of the learning rule on vibrational resonance}
The effect of the learning rate $k_\omega$ on vibrational resonance is illustrated in Fig.~8. As shown in Figs.~8(a) and 8(c), increasing $k_\omega$ leads to a nonlinear decrease in the critical value $B_c$ required to induce vibrational resonance. Figure 8(a) also shows that a large value of $k_\omega$ may cause divergent response. In Fig.~8(c), the consistency between the numerical results of $B_c$ and the corresponding approximated theoretical predications solved by Eq.~(11) is verified once again. In addition, Fig.~8(d) demonstrates that as $k_\omega$ increasing, the maximal response amplitude $Q_{\max}$ decreases, with $Q_{\max}$ exhibiting rapid fluctuations as $k_\omega$ varies. This trend is indirectly reflected in Fig.~8(b) as well. Therefore, excessively large values of $k_\omega$ are not recommended for achieving strong vibrational resonance output.
\begin{figure}
\center
\includegraphics[width=0.5 \textwidth]{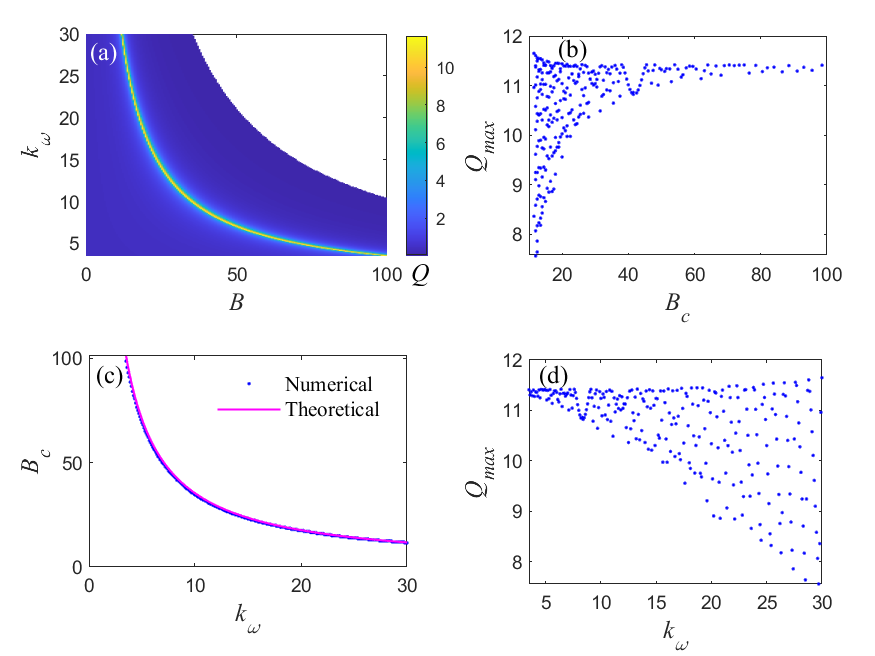}
\caption{Effect of the learning rate $k_\omega$ on vibrational resonance that induced by the auxiliary signal. (a) The response amplitude $Q$ presents obvious vibrational resonance band in the $B-k_\omega$ plane. (b) The maximal value of the response amplitude $Q_{max}$ versus the corresponding critical value $B_c$. (c) The critical values of $B_c$ versus the learning rate $k_\omega$ are obtained by numerical simulation and approximated theoretical predication in Eq.~(11) respectively. (d) The maximal value of the response amplitude $Q_{max}$ versus the learning rate $k_\omega$. The simulation parameters are $\zeta=0.1$, $b=1$, $\beta=10$, $A=0.1$, $\Omega_1=5$ and $\Omega_2=50$.}
\end{figure}
\\
\indent In Fig.~9, the effect of the learning rate $k_\omega$ on vibrational resonance is examined with $\Omega_2$ as the control variable. Figure 9(a) shows that a higher auxiliary frequency $\Omega_2$ can suppress the divergence to some extent, whereas large values of $k_\omega$ may promote divergence. As shown in Figs.~9(a) and 9(c), increasing $k_\omega$ raises the critical frequency $\Omega_{2c}$ at which vibrational resonance occurs. $\Omega_{2c}$ and $k_\omega$ exhibit an approximately relationship expressed by Eq.~(12), which is indicated by the good agreement of the numerical results with the approximated predications of $\Omega_{2c}$. In Fig.~9(d), as $k_\omega$ increasing, the fluctuation in the maximum response amplitude $Q_{\max}$ with respect to $k_\omega$ diminishes and gradually stabilizes at a higher value.
\begin{figure}
\center
\includegraphics[width=0.5 \textwidth]{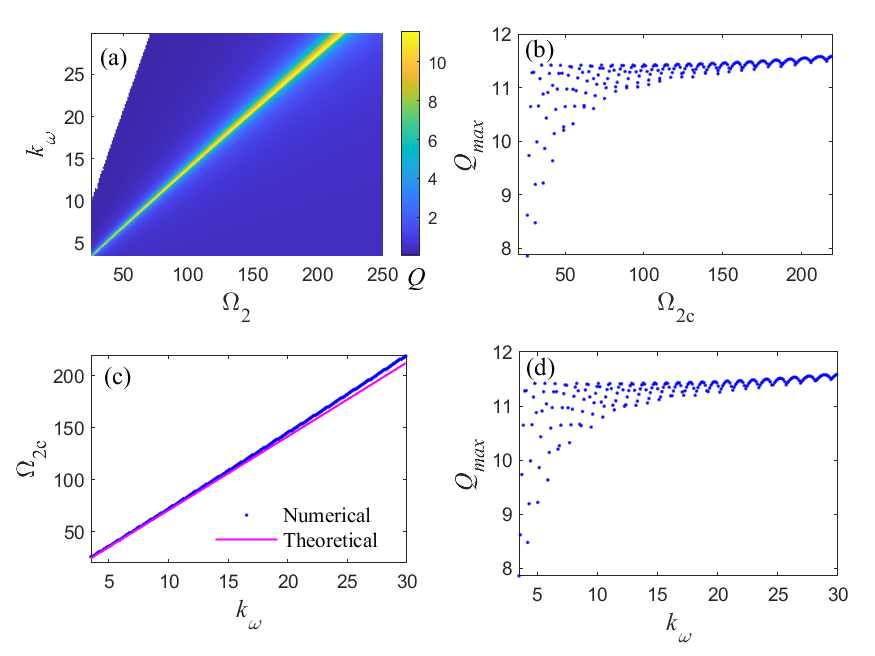}
\caption{Effect of the learning rate $k_\omega$ on vibrational resonance that induced by the auxiliary signal. (a) The response amplitude $Q$ presents obvious vibrational resonance band in the $\Omega_2-k_\omega$ plane. (b) The maximal value of the response amplitude $Q_{max}$ versus the corresponding critical value $\Omega_{2c}$. (c) The critical values of $\Omega_{2c}$ versus the learning rate $k_\omega$ are obtained by numerical simulation and approximated theoretical predication in Eq.~(12) respectively. (d) The maximal value of the response amplitude $Q_{max}$ versus the learning rate $k_\omega$. The simulation parameters are $\zeta=0.1$, $b=1$, $\beta=10$, $A=0.1$, $\Omega_1=5$ and $B=50$.}
\end{figure}
\\
\indent In Fig.~10, the combined effects of nonlinear stiffness $b$ and learning rate $k_\omega$ on vibrational resonance performance are analyzed. When $b$ varies within a small range, its influence on the critical learning rate $k_{\omega c}$ required for resonance is minimal. However, when $b$ changes from negative to positive over a wider range, the critical learning rate $k_{\omega c}$ decreases. The trend in Fig.~10 is consistent with that observed in Fig.~6, although Fig.~6 uses $B$ as the control parameter, whereas Fig.~10 uses $k_\omega$. Figure 10(a) further shows that when $b$ is negative and $k_\omega$ is small, the system response may diverge. Therefore, it is advisable to select a positive value of $b$ corresponding to a hardening spring. At the same time, choosing a smaller positive value of $b$ allows for a larger $Q_{\max}$, thereby ensuring improved vibrational resonance performance.
\begin{figure}
\center
\includegraphics[width=0.5 \textwidth]{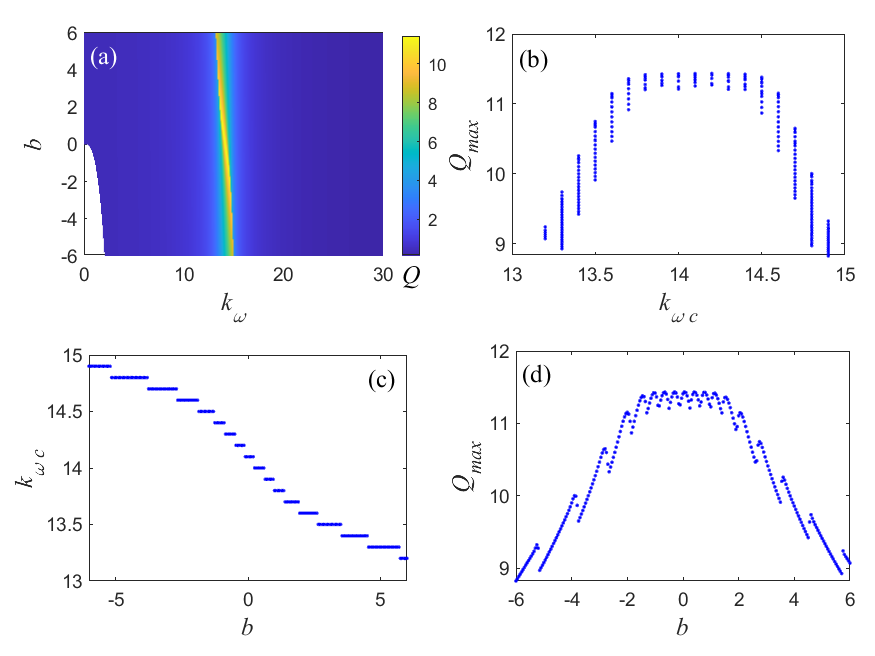}
\caption{Effect of the nonlinear stiffness coefficient $b$ on vibrational resonance that induced by the learning rate $k_\omega$. (a) The response amplitude $Q$ presents obvious vibrational resonance band in the $k_\omega - b$ plane. (b) The maximal value of the response amplitude $Q_{max}$ versus the corresponding critical value $k_{\omega c}$. (c) Critical value $k_{\omega c}$ versus the nonlinear stiffness coefficient $b$. (d) The maximal value of the response amplitude $Q_{max}$ versus the nonlinear stiffness coefficient $b$. The simulation parameters are $\zeta=0.1$, $\beta=10$, $A=0.1$, $\Omega_1=5$, $B=25$ and $\Omega_2=50$.}
\end{figure}
\\
\indent Figure 11 examines the adaptability of the learning rate $k_\omega$ to the characteristic frequency $\Omega_1$. As $\Omega_1$ increasing, it needs a larger learning rate $k_{\omega c}$ to induce vibrational resonance. Apparently, the resonance peak is strongly influenced by $k_\omega$. Moreover, the maximal response amplitude $Q_{\max}$ exhibits larger fluctuations as $k_\omega$ increasing. Therefore, at high characteristic frequencies, it is important to consider multiple parameters in combination to prevent divergence of the system response and to achieve optimal vibrational resonance performance. In addition, the agreement of the numerical results with approximated theoretical predications in the $\Omega_{1c}-k_\omega$ curve validates the effectiveness of the resonance condition given in Eq.~(14).
\begin{figure}
\center
\includegraphics[width=0.5 \textwidth]{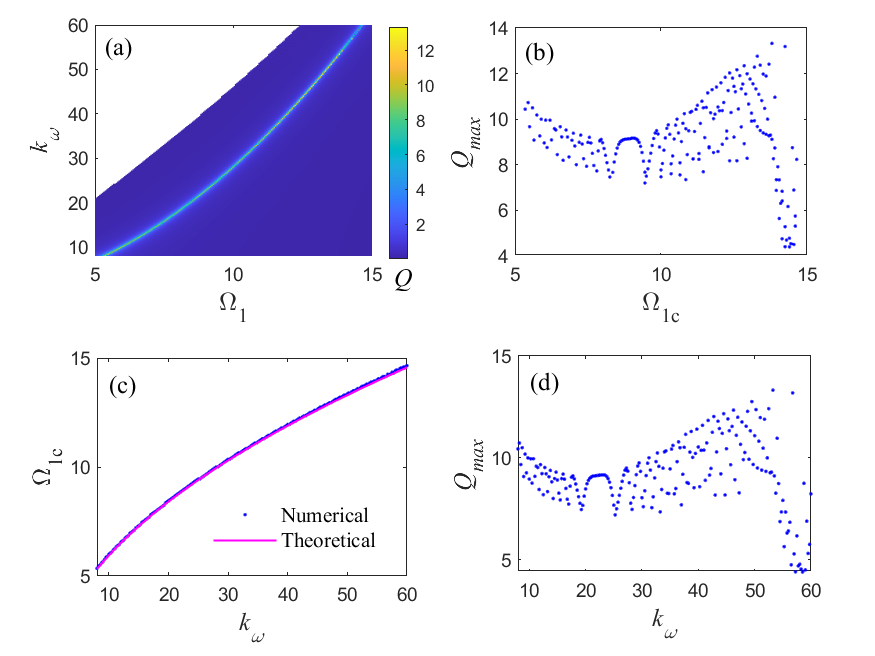}
\caption{Effect of the learning rate $k_\omega$ on vibrational resonance that induced by the characteristic signal. (a) The response amplitude $Q$ presents obvious vibrational resonance band in the $\Omega_1-k_\omega$ plane. (b) The maximal value of the response amplitude $Q_{max}$ versus the corresponding critical value $\Omega_{1c}$. (c) The critical values of $\Omega_{1c}$ versus the learning rate $k_\omega$ are obtained by numerical simulation and approximated theoretical predication in Eq.~(14) respectively. (d) The maximal value of the response amplitude $Q_{max}$ versus the learning rate $k_\omega$. The simulation parameters are $\zeta=0.1$, $b=1$, $\beta=10$, $A=0.1$, $B=50$ and $\Omega_2=10\Omega_1$.}
\end{figure}
\section{Circuit simulation}
Simulating the output of the circuit is an effective way to verify the numerical simulation results. This processing method has been extensively adopted in literature. This section uses circuit to simulate the response of the frequency-adaptive Duffing system, and to verify the occurrence of vibrational resonance. Corresponding to the frequency-adaptive Duffing system in Eq.~(6), we built a circuit by the Multisim software, as shown in Fig.~12.
\begin{figure}[h]
\center
\includegraphics[width=0.60 \textwidth]{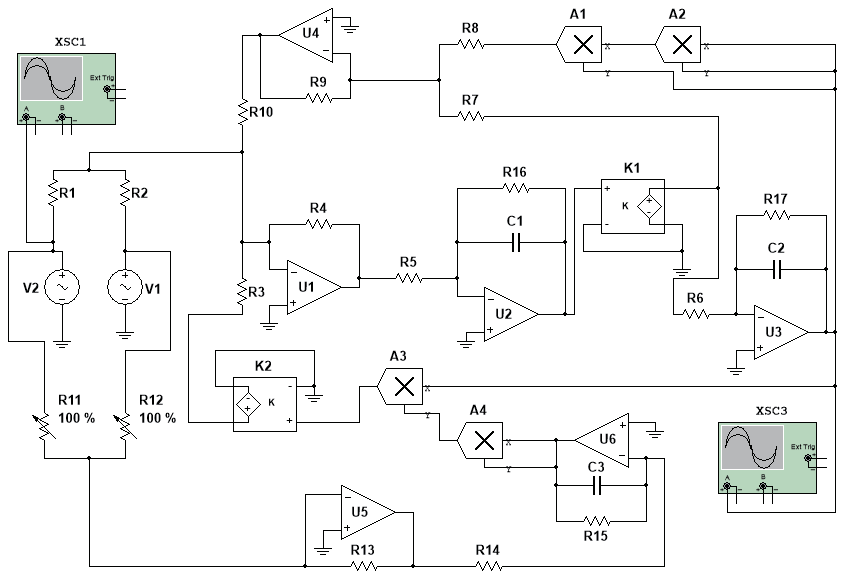}
\vspace{-30pt}
\caption{The circuit of the frequency-adaptive learning Duffing system corresponding to Eq.~(6).}
\end{figure}
\\
\indent The differential equation of the circuit of Fig.~12 is
\begin{equation}
\left\{ \begin{array}{l}
 {R_6}{C_2}\frac{{{d^2}x}}{{d{t^2}}} = {K_1}{R_5}{C_1}\frac{{{R_9}}}{{{R_7}}}\frac{{{R_4}}}{{{R_{10}}}}\frac{{dx}}{{dt}} + {K_2}\frac{{{R_4}}}{{{R_3}}}{\omega ^2}x - \frac{{{R_9}}}{{{R_8}}}\frac{{{R_4}}}{{{R_{10}}}}{x^3} + \frac{{{R_4}}}{{{R_1}}}{f_1}\left( t \right) + \frac{{{R_4}}}{{{R_2}}}{f_2}\left( t \right) \\
 {R_{14}}{C_3}\frac{{d\omega }}{{dt}} = \frac{{{R_{13}}}}{{{R_{11}}}}{f_1}\left( t \right) + \frac{{{R_{13}}}}{{{R_{12}}}}{f_2}\left( t \right) \\
 \end{array} \right.
\end{equation}
where ${f_1}\left( t \right) = A\cos \left( {{\Omega _1}t} \right)$ and ${f_2}\left( t \right) = B\cos \left( {{\Omega _2}t} \right)$, which are represented by $V_1$ and $V_2$ in Fig.~12. The component parameters of the circuit are as follows: ${R_1} = {R_2} = 1k\Omega $, ${R_3} = {R_4} = {R_8} = {R_9} = {R_{10}} = {R_{13}} = 10k\Omega$, ${R_5} = {R_6} = {R_{14}} = 200k\Omega$, ${R_7} = 50k\Omega$, ${R_{15}} = {R_{16}} = {R_{17}} = 1000 k\Omega$, ${R_{11}}$ and ${R_{12}}$ are adjustable resistors used to adjust the parameter $k_\omega$ and achieve adaptive control of the resonance. In addition, ${R_{15}}$, ${R_{16}}$ and ${R_{17}}$ are used to enhance the stability of the amplification process and improve distorted waveforms. The capacitance values are ${C_1} = {C_2} = {C_3} = 5\mu F$. A1, A2, A3, and A4 are multipliers used to perform multiplication operations on two signals X and Y. In practical circuits, to ensure the amplification of the signal amplitude, suitable operational amplifiers are needed. In this study, PA83 is recommended. K1 and K2 are voltage gain modules with an amplification factor $K=-1$.

\indent Based on the constructed frequency-adaptive learning Duffing circuit, herein, we compare the output of the circuit and the numerical results in Sec.3. As a specific example, we refer the example in Fig.~3. For example, taking $\Omega_2=40$ in Fig.~3, the numerical critical value for vibrational resonance is $B_c=27.6$ and the corresponding peak value is $Q_{max}=11.3031$, as given in Fig.~3(d). In addition, the approximated theoretical predication of $B$ is $B_c=28.273$ for this case, as shown in Fig.~3(c). Under all identical simulation parameters, the input signal is shown in Fig.~13(a) of the virtual oscilloscope. Herein, $100mV$ indicate the amplitude of the input signal is $0.1V$. In Fig.~13(b), the amplitude of the input signal is $B=B_c=27.6$, we get the waveform of the output that indicating the appearing of vibrational resonance. As shown in the virtual oscilloscope, after passing through the designed circuit system, the amplitude of the characteristic signal is amplified by about 8 times. It can be obtained directly from the virtual oscilloscope, i.e., the two values marked in the virtual oscilloscope boxes. These two values are labelled at two time points at the valley and peak of the waveform respectively. In Fig.~13(c), similarly, corresponding to the approximated theoretical value $B_c=28.273$, the amplitude of the input signal is amplified by about 10 times. Both from the two output results of the circuit, the effectiveness of the numerical results and the significance of the frequency-adaptive learning Duffing system are validated simultaneously.
\begin{figure}
\includegraphics[width=0.75 \textwidth]{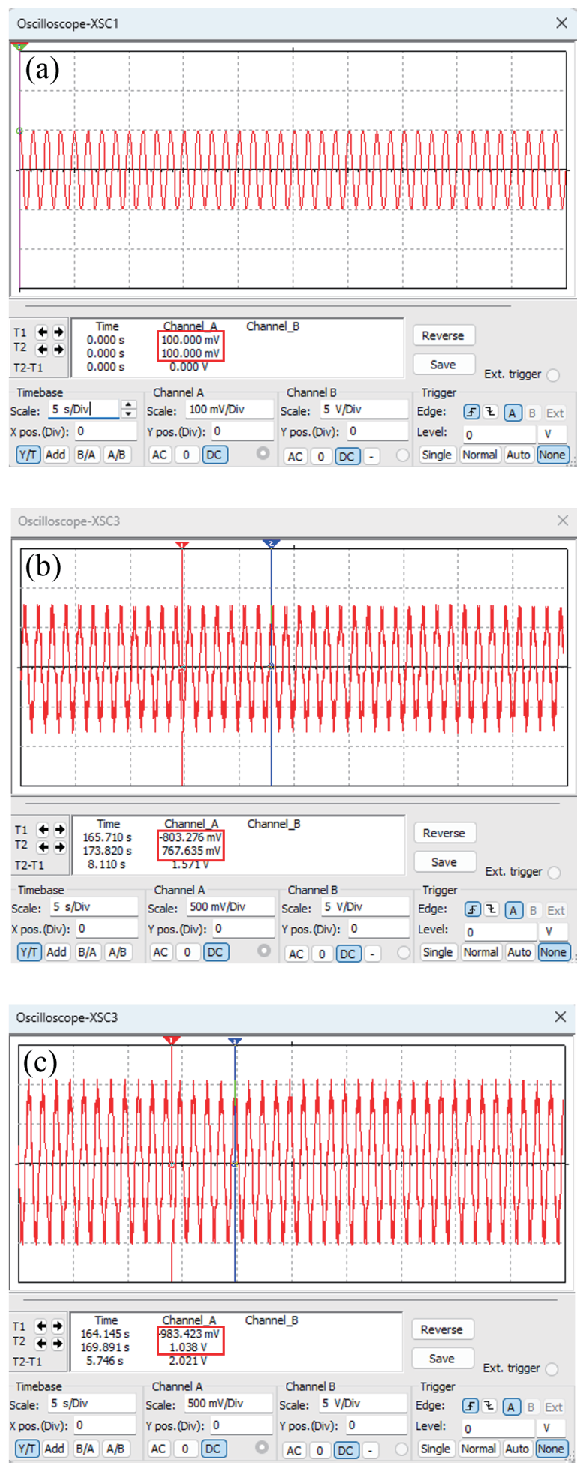}
\caption{The vibrational resonance is realized by the simulation circuit. (a) The waveform of the input weak signal. (b) The waveform of the output corresponding to vibrational resonance occurs at the numerical critical value $B_c=27.6$. (c) The waveform of the output corresponding to vibrational resonance occurs at the approximated theoretical critical value $B_c=28.273$.}
\end{figure}
\section{Discussions}
\subsection{Comparison to other frequency-adaptive learning rules on vibrational resonance}
\indent The frequency-adaptive learning rule in Eq.~(6) can take other different forms. Herein, we let $f(t)= A\cos ({\Omega _1}t) + B\cos ({\Omega _2}t)$ for the convenience of describing. In addition to the frequency-adaptive learning rule that we discussed in Sec.3, we consider two alternatives: $\dot \omega  = {k_\omega }xf(t)$ and $\dot \omega  = {k_\omega }\dot xf(t)$. These two additional rules can be regarded as the so-called Hebbian learning rules. Some works on Dynamic Hebbian learning in different frequency-adaptive oscillators have been reported \cite{Ref34, Ref35, Ref41, Ref42, Ref43}. Figure 14 compares the vibrational resonance performance under all three learning rules mentioned here, using the same simulation parameters. In each case, the response amplitude can achieve a large value, indicating that all three frequency-adaptive learning rules are capable of inducing vibrational resonance. Among them, the rule $\dot{\omega} = k_\omega f(t)$ produces the strongest vibrational resonance, with the largest $Q_{\max}$, as can be seen by comparing Fig.~14(a) with Figs.~14(b) and 14(c). For the two frequency-adaptive alternative learning rules, $\dot{\omega} = k_\omega x f(t)$ and $\dot{\omega} = k_\omega \dot{x} f(t)$, vibrational resonance can occur even at small values of $B$, but the response is also more prone to divergence, as illustrated in Figs.~14(b) and 14(c). In contrast, the learning rule $\dot{\omega} = k_\omega f(t)$ demonstrates better performance: it produces a larger resonance peak, achieves resonance at reasonable values of $B$, and, most importantly, yields a response that is less susceptible to divergence. This characteristic makes it particularly advantageous in weak signal processing applications.
\begin{figure}
\center
\includegraphics[width=0.5 \textwidth]{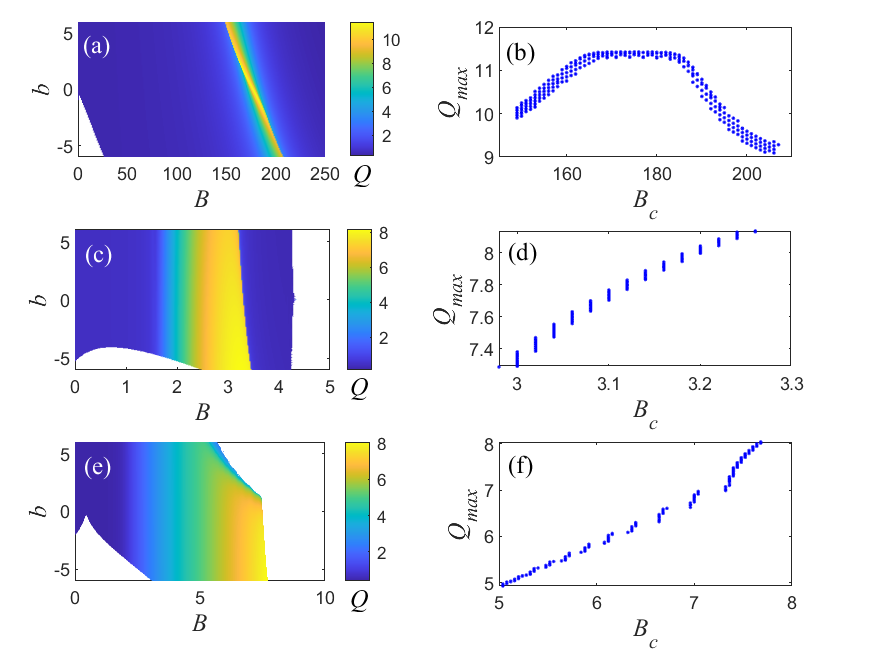}
\caption{Vibrational resonance induced by different learning rules. (a) $\dot \omega  = {k_\omega }f(t)$, (b) $\dot \omega  = {k_\omega }xf(t)$, (c) $\dot \omega  = {k_\omega }\dot xf(t)$. The simulation parameters are $\zeta=0.1$, $\beta=10$, $A=0.1$, $\Omega_1=5$, $\Omega_2=50$, $k_\omega=2$.}
\end{figure}
\subsection{Superiority and limitation of the frequency-adaptive learning Duffing system}
From the above research, we have listed some superiorities of the new frequency-adaptive learning Duffing system. Compared with two other commonly used Hebbian learning rules, the proposed frequency-adaptive learning rule has better performances on the stability and resonance degree although the new rule is simple. Significantly, the characteristic frequency of a weak periodic signal that cause vibrational resonance in the system is greatly broadened in the frequency-adaptive system. The numerical results have also been verified by simulation circuit which provides reference value in hardware applications in the future.\\
\indent Although some superiorities of the frequency-adaptive learning Duffing system on vibrational resonance are discussed in detail, the work is insufficient in some aspects due to the length and reasonable structural arrangement of the paper. It involves both the limitation of the system, the insufficient of the study, and the prospects of the research. First, we must ensure the stability of the system to use it in weak signal processing. From the results of Sec.3, we find that there are unstable regions in some parameter ranges. From the perspective of signal processing, we can get a satisfactory signal processing effect only in parameters with with stable response. As a result, the stability of the frequency-adaptive learning Duffing system is one focus of the future works. Moreover, the theoretical results, such as the resonance conditions in Eqs.(10)-(14), are very approximated results and some important parameters are not contained in them, such as the parameters $\zeta$, $b$ and $\beta$. Adopting some more precise analytical methods for theoretical analysis is also one of the future important tasks. Further, vibrational resonance focus on a pure weak signal. However, the weak signals are usually contained many frequency components and often drowned out by strong noise in real engineering sites. Attempting to use the vibrational resonance of a specific frequency-adaptive learning nonlinear system to process weak and complicated signals in strong noise background is also an interesting and worthwhile topic for specialized research.
\section{Conclusions}
This work investigates the vibrational resonance phenomenon in a frequency-adaptive learning Duffing system. In the conventional monostable Duffing oscillator, vibrational resonance is difficult to achieve when the system is excited by a high-frequency characteristic signal. When both the characteristic and auxiliary signals are simultaneously amplified and applied, vibrational resonance can occur in the case of a hardening spring, but not in the case of a softening spring, where the system response tends to diverge.

To address this limitation, a new frequency-adaptive learning Duffing system is proposed. By introducing a simple frequency-adaptive learning rule, the system achieves reliable vibrational resonance performance, even under conditions where the conventional Duffing system fails. Moreover, a detailed analysis of the effects of signal parameters, system parameters, and the learning rule is carried out, with particular emphasis on the relationship between the vibrational resonance peak and control parameters. The numerical simulation results are verified by approximated theoretical predications and circuit simulations. Based on these findings, guidelines for designing a learning nonlinear system to achieve optimal vibrational resonance are provided from different aspects.

Finally, distinction of three different learning rules on vibrational resonance are compared, including two Hebbian learning rules that have been widely used in prior studies. The results show that the proposed frequency-adaptive learning rule is not only simpler in the form but also improves the stability of the system, prevents the response divergence, and yields a stronger vibrational resonance output. Its performance is significantly better than that of the other two frequency-adaptive learning rules. These results highlight the potential value of the frequency-adaptive learning Duffing system for processing weak characteristic even much more complicated signals, and they also suggest promising ways for related engineering applications in the future.
\section*{CRediT authorship contribution statement}
{\bf Zhongqiu Wang}: Conceptualization (equal); Formal analysis (lead); Methodology (supporting); Writing - original draft (lead). {\bf Jianhua Yang}: Conceptualization (equal); Data curation (equal); Funding acquisition (equal); Methodology (lead); Writing - review \& editing (equal). {\bf Feng Tian}: Data curation (equal); Software (lead). {\bf Huatao Chen}: Data curation (supporting); Formal analysis (supporting). {\bf Miguel A. F. Sanju\'an}: Conceptualization (equal); Funding acquisition (equal); Writing - review \& editing (equal).
\section*{Declaration of competing interest}
The authors declare that they have no known competing financial interests or personal relationships that could have appeared to influence the work reported in this paper.
\section*{Acknowledgments}
The project was supported by the National Natural Science Foundation of China (Grant No. 12472036), and the Spanish State Research Agency (AEI) and the European Regional Development Fund (ERDF, EU) under Project No. PID2023-148160NB-I00. We appreciate the support provided by Haibo Yu in the circuit simulation. We also thank the anonymous reviewers for their constructive feedback on the paper revisions.
\section*{Data Availability Statement}
The data that support the findings of this study are available from the corresponding author upon reasonable request.
\end{document}